\documentclass[a4paper,11pt]{article}
\pdfoutput=1 

\usepackage[T1]{fontenc} 

\usepackage{amssymb}
\usepackage{amsthm}

\usepackage{mathtools}
\usepackage{amsfonts}

\usepackage{subcaption}

\usepackage{hyperref}

\newcommand{\be}{\begin{equation}}
\newcommand{\ee}{\end{equation}}

\newcommand{\bi}{\begin{itemize}}
    \newcommand{\ei}{\end{itemize}}

\title{Gauss-Manin equations for propagators in the case of arbitrary masses}

\author{ S. Srednyak, Duke University, Durham, USA}

\begin{document}

\maketitle

\abstract{
We derive the complete list of singularities of propagators in theories with arbitrary (complex) masses and for arbitrary diagram. We derive in a closed form differential equations for the propagator as a function of the momentum and masses.
}

\tableofcontents

\section{Introduction}

In this paper we derive the Gauss-Manin connection for propagators for arbitrary mass case. Differential equations for perturbative amplitudes play an important role in their theory. They can be used for numerical evaluation of the amplitudes as well as for analytic study. There has been a large interest in this topic for the case of propagators. In particular, the papers  \cite{Adams:2013nia,Remiddi:2013joa,Bloch:2016izu} address the case of sunrise graphs. More recently, there is more work on the banana family \cite{Klemm:2019dbm,Pogel:2022vat}. Beginning steps of all these analyses is the derivation of differential equations for the function represented by the diagram.

We observe that singularities of propagators can be completely analysed in closed form. In particular, we derive explicit equations for all of the singularities of the diagram function , relating them to the combinatorics of the diagram. Our derivation is based on the observation that in appropriate coordinates the Landau polynomials are linear functions of the coordinates. This result is non tivial because it is known \cite{lee2015reducing} that systems with regular singularities can possess higher order poles and apparent singulartities. We use tools from the theory of hypergeometric functions and holonomic D-modules to obtrain our results.

\section{Aknowledgemets}

The author was supported in part by  BNL LDRD 21-045S NPP.

\section{Preliminaries}

We consider the standard propagators in massive theories \cite{bogner2010feynman} that can be written in the form
\be
J_m(p,m_i) = \int \frac{1}{\prod ((q_i+\delta_ip)^2+m_i^2)} q^m \prod_{a=1}^L d^{d}q_a
\label{eq:integral}
\ee
where $\delta_i= 0,1$, in general depending on the choice of momentum flow. Here 
\be 
q_i = \sum_{a=1}^L l_{i,a}q_a 
\ee
\be 
l_{i,a} = 0,\pm 1
\ee 
$q_a$ is a basis of loop momenta and $l_{i,a}$ are combinatorial coefficients defined by momentum flow.  In particular, this family includes the unequal mass sunrise intgrals considered in \cite{Bloch:2016izu}. These integrals have the form 
\be 
I = \int d^d q_1 d^d q_2 \frac{1}{q_1^2+m_1^2} \frac{1}{q_2^2+m_2^2} \frac{1}{(q_1+q_2+p)^2+m_3^2}
\ee

\section{Results}

In this section we formulate our main results.

{\it
{\bf Th1} (Characterization of singularity locus of the propagator). The singularity locus of the propagator is given by the set
\be
x  = r_{i_1} m_{i_1} + r_{i_2} m_{i_2} + .... + r_{i_s} m_{i_s}
\ee
where $r_{i_s}$ are rational numbers that depend on the diagram. There is finite set of such numbers, the cardinality of which grows exponentially with the number of loops.
\qedsymbol{}
}

{\it
{\bf Prop.} The singularities of the propagator in masses are located at the set
\be
\sum r'^D_i m_i  =0
\ee
for some integer numbers $r'^D_i$ that depend on the diagram .

}

{\it
{\bf Th2} ( Differential equations for the propagator). The propagator satisfies the following system of equations
\be
\frac{\partial f^D}{\partial z_k } = ( \sum_{R=\{r_{i_1}...r_{i_s}\}}  \frac{ A^D_{k,R}}{ x + r_{i_1} m_{i_1} + r_{i_2} m_{i_2} ...  r_{i_s}m_{i_s}}   )f^D
\ee
where the sum is extended over the set described in Th1. Here $D$ denotes a diagram and $z_k = \{\sqrt{p^2,m_1,...,m_I} \}$ is any of the variables on which the diagram function depends. The matrices $A^D_{k,I}$ depend only on coupling, dimension and diagram topology, and have no dependence on masses or the momentum.  

Analogous formula holds for the sum over all diagrams.
\qedsymbol{}
}

\section{Proofs}

\subsection{Proof of Th1}

We will carry out the proof only for leading singularities. The analysis of subleading singularities reduces to the case of sub diagrams. For leading singularity, there is a set of propagators $D_{i_s}(p,q_i)$ that develop a vanishing cycle in their intersection
\be
D_I = \cap_{i_s \in I} D_{i_s}
\ee
We will consider the vanishing cycles at finite distance in q-space. The emergence of vanishing cycles corresponds to degeneracy of the system of normals to the varieties $\{q: D_i(p,q) =0\}$. Note that these varieties are highly degenerate. Each of them has as singularity locus a $(L-1)d$-dimensional plane in q-space.\footnote{This is most easily seen in an example. The quadric $x_3^2+x_4^2=0$ in $\mathbb{C}^4$ with coordinates $x_1,x_2,x_3,x_4$ has the plane $(x_1,x_2,0,0)$ as singularity.} Nonetheless, our criterion still works. It can be seen by applying the general theory of vanishing cycles as applied to singular schemes \cite{kashiwara1985microlocal,ginsburg1986characteristic}.

The criterion formulated above results in equations
\be
\sum b_i ( \delta_{s,i} p +q_{s,i,1}+...+q_{s,i,r_i}) =0,\  s \in \{0,...,L\}
\ee
for some $b_i$ that are not all zero. These equations can be solved for $q_i$ as
\be
q_a = \alpha_a p
\ee
fo certain coefficients $\alpha_a$. The singularities develop only when all vectors are collinear. Then the equations $D_s(p,q) =0$ can be rewritten in the form
\be
D_s = (\delta_{s}p + q_{s,1}+...q_{s,r_s})^2 -m_s^2 = (\delta_s+a_{s,1}+...+a_{s,r_s})^2 p^2 -m_s^2 = 0
\ee
or
\be  
\delta_s+a_{s,1}+...+a_{s,r_s} = \pm m_s/x
\ee
After elimination of the variables $a_{s,i}$ we obtain our claim.

As a byproduct of our proof, we obtain the following conditon on the singularities 

\begin{equation}
\begin{vmatrix}
    l_{i_1,1} & l_{i_1,2} & ... & l_{i_1,L} & 1-\frac{m_{i_1}}{x} \\
    l_{i_1,1} & l_{i_1,2} & ... & l_{i_1,L} & 1-\frac{m_{i_1}}{x} \\
    ... \\
    l_{i_{L+1},1} & l_{i_{L+1},2} & ... & l_{i_{L+1},L} & 1-\frac{m_{i_{L+1}}}{x} \\
\end{vmatrix}
=0
\end{equation}

The case of the vanishing cycle at infinity is simpler. In this case, we simply drop the terms $m_i^2$ and consider $q_i$ as homogeneous coordinates on the projective space $\mathbb{CP}^{Ld-1}$. Considerations similar to the above lead to the equation
\be
p^2 =0
\ee
which is the only leading Landau singularity at infinity.

\subsection{Apparent singularities and non Fuchsian regular singular points}

In this subsection we discuss Th2 from the point of view of general theory of regular singularities. The existence of Gauss-Manin connection for the periods of rational formas was established in \cite{griffiths1968periods}. The block structure of the connection is discussed in \cite{schmid1973variation}.

The statement of our theorem does not follow from general principles concerning falt bundles with a given set of regular singular points because:

\bi 
\item There can be higher order poles that nonetheless lead to regular solutions. 

\item There can be apparent singularities.

\ei

Case 1. is documented in \cite{anosov2013riemann},chapter 2. The study of reduction of order of poles by algebraic gauge transformations can be very fruitful and has lead to computational algorithms, see \cite{moser1959order,barkatou1997algorithm,barkatou2009moser}.

Case 2. is subject to numerous works \cite{lee2015reducing,v2006additional,v2007irreducible,gontsov2004refined,gontsov2008solutions,shiga2004triangle,cohen2002application}. It has interesting relation to isomonodromy \cite{haraoka2007middle,its2018monodromy} and to transcendental number theory \cite{shiga2004triangle,cohen2002application}.

These examples preclude the conclusion of our statement from general principles and necessitate the use of alagebraic methods.

From general prrinciples, we could only conclude the following form of the connection 
\be 
\frac{dJ}{dx} = \sum_k \sum_{I_1,...,I_k} \frac{A}{\prod_I (x-x_I)^n_I} J 
\ee 
where $x_I$ include the singularities given by the vanishing cycle condition and perhaps some additional singularities. Some of the singularities resulting from vanishing cycle conditon can actually be apparent singularities. The decision process that would allow to find out which of these singularities are apparent proceeds through the analysis of action of the intersection pairing matrix for middle dimensional cycles on the hypersurface $\{q|P(q)=0\}$ on the original cycle of integration and determining which elements of the orbit of such action correspond to long cycles at each of the components of vanishing loci.

\subsection{Proof of Th2}

To prove our theorem, we embed the integral  \ref{eq:integral} into a family of integrals, generically deforming its coefficients ( versal deformation in terminology of \cite{arnold}). I.e. we consider the integral 
\be 
I(L) = \int_\Delta (L(x))^a D(x)^b d^dx
\ee 
where the coefficients of the polynomial $L$ are now genric complex numbers. For such integrals, the existence of Gauss-Manin connection was established in \cite{griffiths1968periods}. An alternative, constructive approach can be obtained as follows. First, according to \cite{gelfand1990generalized} there exists holonomic D-module that this hyperfunction $I(L)$ satisfies. It is given, constructively, by the ideal generated by
\begin{gather}
    \sum \omega_\mu L_\omega \frac{\partial}{\partial L_\omega} - \beta_\mu  \\ 
    \prod_{\omega \in \Omega'} \frac{\partial^{a_\omega}}{\partial a_\omega} - \prod_{\omega \in \Omega''} \frac{\partial^{a_\omega}}{\partial a_\omega}
\end{gather}
This holonomic D-module is equivalent, on the generic stratum, to the Gauss-Manin connection ( see \cite{saito2013grobner,hibi2017pfaffian}). This equivalence is obtained by the choice of basis of itegrals. One such choice is provided by $I_m(L) = \int_\Delta (L(x))^a D(x)^b x^m d^dx$. The Gauss-Manin connection can be written as 
\be 
\frac{\partial}{\partial L_\omega}I_m(L) = \sum_{i} \frac{A_{\omega,i,m,m'}(L)}{D_i(L)} I_{m'}(L)
\ee
where $D_i(L)$ are equations for compoenents of the discriminantal locus of $L$.

Our theorem then follows by restricting the Gauss-Manin connection to 1-dimensional subspace of the deormed parmeter space $\{L\}$ \cite{saito2013grobner}.

\section{Discussion}

Our result fits propagators of quantum field theories into the family of equations
\be
\frac{ \partial f(x,l)}{\partial x_k } = ( \sum \frac{A_{k,I}}{l_I} ) f
\ee
where $A_{k,I}$ are constant matrices and $l_{I}$ are functions linear in the variables $x_i$. This family of functions provides a natural generalization of the Grassmannian hypergeometric functions considered in \cite{aomoto2011theory,gelfand1982geometry}. It is desirable to obtain combinatorial characterization of their solutions. To solve this problem, it is necessary to consider their dependence on the parameters $l_{I,i}$. This dependence leads to the study of irregular singularities \cite{jimbo1981monodromy,deift1999algebro}. Full solution of this problem would involve quantum groups  \cite{varchenko1995multidimensional,xu2019closure}

\section{Comparison with the known literature}

In papers \cite{Bloch:2016izu, Adams:2013nia} deep properties of the sunrise family were studied. Sunrise graphs fall inside the class of the diagrams that we consider. While the equations derived in \cite{Bloch:2016izu, Adams:2013nia} are seemingly more complicated, they in fact were derived before in \cite{Caffo:1998du}. The examination of formulas (7) and (12) of \cite{Caffo:1998du} shows that the equations of \cite{Caffo:1998du} can be transformed into the form stated in our theorem . The polynomial D in this paper can be decomposed into a product of linear forms, after which a partial fractioning procedure can be applied.

Our method can be applied to the family of banana graphs \cite{Pogel:2022vat,Klemm:2019dbm}.

\section{Conclusion}

In this paper we obtained Gauss-Manin connection for propagators that depend on an arbitrary set of masses. We hope our results can be useful for numerical computation of propagators and self energies.

\bibliographystyle{unsrt}
\bibliography{ref_gauss}

\end{document}